\documentclass{article}
\usepackage{graphicx}
\begin{document}
\begin{flushleft}

\end{flushleft}
\begin{center}
\begin{Large}
\textbf{Photomapping Using\\ Aerial Vehicle in USARSim}\\
Su Kim

\end{Large}
\end{center}
\begin{abstract}
 Creating a photomap plays a critical role in navigation. Therefore, flying vehicles
are usually used to create topdown maps of the environment. In this report we used two
different aerial vehicles to create a map in a simulated environment
\end{abstract}

\section{Introduction to USARSim( Unified System for Automation and Robot Simulation)}

USARSim is a high fidelity simulation of urban search and rescue (USAR) robots and environments intended as a research tool for the study of human-robot interaction (HRI) and multirobot coordination. USARSim is designed as a simulation companion to the National Institute of Standards and Technology (NIST) reference test facility for autonomous mobile robots for urban search and rescue\cite{usarsim}.\\
As the time passed, the USARSim has been used for broader range of robot simulations. The acronym was converted to Unified System for Automation and Robot Simulation. But it is still USARSim.\\

Aerial vehicles in this environment play a very important role\cite{survey, shams2}. They act as the eyes of a bird, flying in the sky and informing the other robots about the environment. The better they take pictures, the better other robots will cooperate. In the other hand two main questions arise. The first one is how to merge these pictures to have a general map. The second question is what to do with this map. In the following parts we will talk about these issues.\\

There are two aerial vehicles implemented in the USARSim, a Blimp named as Passarola and a Quadcopter named as Airobot as shown in Figure 1. Both of these vehicles have the ability to carry a camera to take some pictures. \\

The best and easiest way of using sequence of pictures is to create a photomap from them. This photomap would be very useful for localization purposes. Creating such a map is not easy and takes a long time to develop an algorithm to handle it. There are several image processing algorithms used for registering sequence of images in order to create a photo map. One of these algorithms that uses Fourier Melin Transform has been developed by Dr Bulow is used in our system. Even if there is such an algorithm, we still have some problems.\\

One of the challenging problems in having a photomap is having a decent sequence of images that an image processing algorithm can handle\cite{survey, img}. One of these problems is movements of the camera. According to the movements of camera the image processing algorithm has to do some scaling, rotation or translation processes. The less the changes, the better the final map will be. Intrinsically, Passarola has a huge amount of inertia that makes its movements smoother. That makes it move without abrupt changes in altitude or position. In the better word, its movements along X, Y, and Z axises will be smoother. Thus we chose passarola for carrying the camera.\\

In the following parts of this report we will explain what we have done in detail.\\

\begin{figure}
\begin{center}
\includegraphics[scale=0.65]{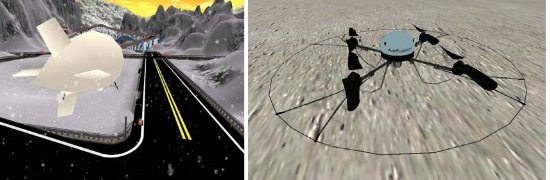} 
\caption{A simulated Passarola on the left and an Airobot on the right side}
\end{center}
\end{figure}

\section{What we have done with the source code}

There is an implementation of FMI in Matlab and another in C++ \cite{shams2, mult}. The C++ one was in the experements folder of Jacobotics source code and was basically developed to work with the Philips webcam in order to test some parts of FMI algorithm. Since that camera was a real camera and there are some issues regarding calibration, they had to be considered during photomaping.  When we are working in a simulated area, no calibration is required\cite{shams3}. The first task we did was to change the FMI source code in C++ language. So we added a new parameter to its constructor under the name of calibration. If it is set in the constructor, calibration will be done. Otherwise it won't do calibration.\\ 

The second task that we did was to move the FMI source code from experimnets to the library folder, and set the pathes and links. Then everybody can use it easily. At the moment it is in the lib/mapping/FMI folder. \\

Then we configured a Blimp to have a downlooking camera and some other things like propellers etc. We added two important sections in Jacobotics source code. One of them is an actuator to drive our blimp, and the other is an image grabber to take pictures and create a photomap of them. The actuator is inside /lib/virtual/actuators/PassarolaDrive. For using this package it is required to create an object from this type and use setMotorSpeeds method. It receives three parameters and returns nothing. This method generally sets the motor speed of one of the propellers. The parameters are xZAngle, thrustPropeller and tailPropeller\\
Where\\ xZAngle, float, is the rotation angle of the support thrust motors bars, that make possible change the
altitude of the robot (i.e up/down). The value is the absolute rotation angle, in radians per second.\\
thrustPropeller, float, is the module of the velocity vector to be applied by the front thrusters, to move the
robot in the X0Z plane (i.e forward/backward and up/down as the value of XZAngle). The value is the absolute linear
velocity, in meters per second.\\
tailPropeller, float, is the rotational velocity (i.e left/right). The value is the absolute rotational velocity, in meters per second.\\

The other section that I added was PassarolaImageGrabber. It is placed in lib/virtual/autonomy/PassarolaImageGrabber. This part is responsible for getting pictures from the image server and store them in files. These files could be used for testing FMI algorithm in offline mode, but it could be easily disabled. Since we are in the phase of testing this algorithm, we prefered to enable this feautre.\\
\begin{figure}
\begin{center}
\includegraphics[scale=0.5]{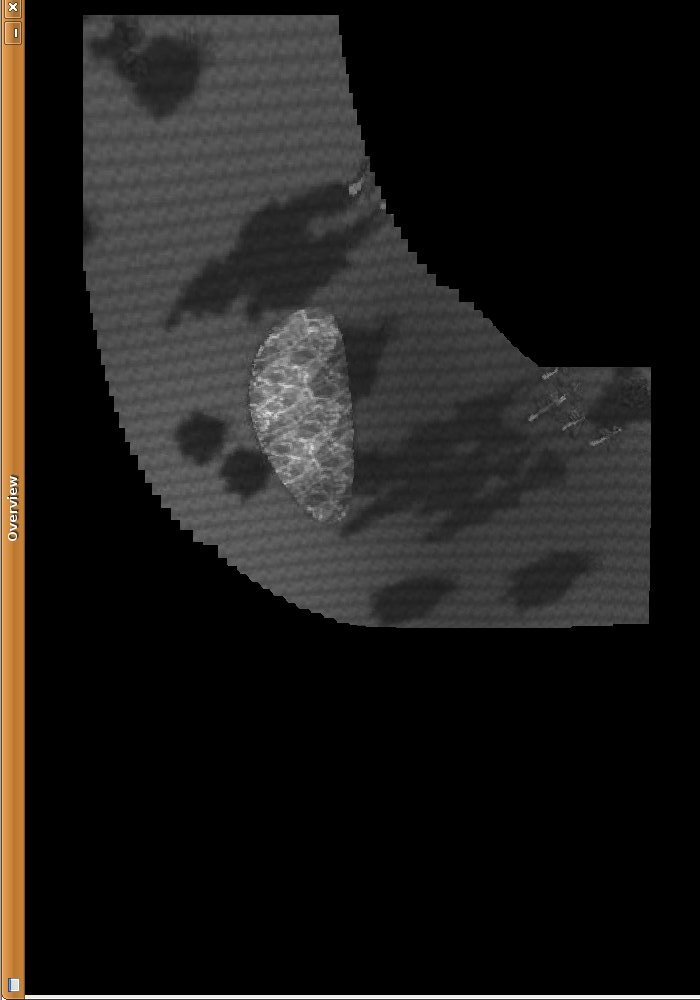} 
\caption{A sample result of photomaping in OpenCV}
\end{center}
\end{figure}

\begin{figure}
\begin{center}
\includegraphics[scale=0.6]{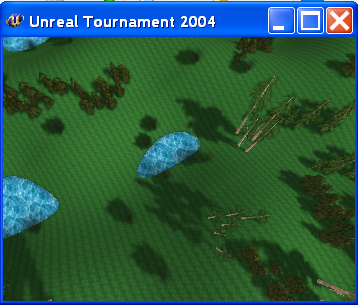} 
\caption{The real environment}
\end{center}
\end{figure}

The next importnat part done in this file is to apply FMI on the sequence of images captured from the camera. Our FMI algorithm works only on special format pictures. Since the image server can take picures only in colored format and rectangular format, we have to convert them to grayscale and square shape. For doing this task we used GraphicMagick library. So for using this part the user has to install this package first. After cropping and converting the picture sequence, we applied FMI algorithm to it. The output of the FMI algorithm is X, Y, Rotation and Scale. These variables associated with the new image can create the whole photomap. In order to create the photomap, we put the new picture in the appropriate place in our GUI. We developed our own GUI using OpenCV to present the output of the algorithm in online mode. One of the outputs is shown in figure 2. The real environment that we applied our algorithm is shown in figure 3. One of the problems of using OpenCV is that it needs a fixed size for the photomap. Since we are creating the map in time, it is a growing map and we can't have a fixed size for it. There are several solutions for handling it, but everything should be done from scratch. So it would be easier to use their GUI.\\

The idea was to send these variables and images to the original GUI . Then we had to send it on Wireless Simulation Server to be captured by that GUI. We talked with them and Ravi said they are to change the Image message and it is not possible to send it to their GUI. The other option is not to use WSS server and use some parts of their source code and develop our own QTObject in order to integrate it with their GUI. But lots of things have to be changed, because their GUI slows down when the number of images is increasing. So it is not again usable and we have to fix this problem of the GUI first. By default QT is repainting everything on its canvas. It doesn't care which part has recetly been changed. This will force high workload on CPU and memory of the computer, and slows down the overal progress. The idea is to render and repaint only the newer part in photomap. Since it was suggested very late, a few days ago, we haven't had enough time to do it yet. But we are trying to do it as soon as possible.\\

\section{Conclusion}

Having a general map of a disaster situation is very useful in a cooperative multirobot system. In this report we explained our approach for integrating FMI algorithm into Jacobotics source code, in order to have a growing map of the environmet in a disaster situation. We showed that the job was done pretty well, but there are still some tasks to be done in near future. The most important task is to take it to a QT GUI from an OpenCV one.\\

\end{document}